\documentclass[12pt]{article}
\usepackage{amsmath,amsfonts,amssymb,amsthm}
\usepackage{mathrsfs}
\usepackage{verbatim}
\usepackage{complexity}
\usepackage{epsfig}
\usepackage{picinpar}
\usepackage{fullpage}

\newtheorem{theorem}{Theorem}[section]
\newtheorem{lemma}[theorem]{Lemma}
\newtheorem{corollary}[theorem]{Corollary}
\newtheorem{claim}[theorem]{Claim}
\newtheorem{proposition}[theorem]{Proposition}
\newtheorem{problem}{Problem}
\newtheorem{observation}[theorem]{Observation}

\newtheorem{question}{Question}

\usepackage{latexsym}
\usepackage{algorithm2e}
\usepackage[all]{xy}

\newcommand{\N}{\mbox{$\mathbb{N}$}}
\newcommand{\F}{\mbox{$\mathbb{F}$}}
\newcommand{\Q}{\mbox{$\mathbb{Q}$}}
\newenvironment{jsay}{\noindent{\bf Jayalal says:}}{\qed\bigskip}

\newcommand{\rank}{\mbox{\sc Rank}}
\newcommand{\code}{\mbox{\sc code}}

\renewcommand{\C}{{\mathcal{C}}}
\newcommand{\B}{{\mathcal{B}}}
\newcommand{\I}{{\mathcal{I}}}
\newcommand{\BB}{{\mathscr{B}}}

\renewcommand{\MI}{\mbox{\sc MI}}

\renewcommand{\GI}{\mbox{\sc GI}}
\newcommand{\PMI}{\mbox{\sc PMI}}

\renewcommand{\GA}{\mbox{\sc GA}}

\renewcommand{\MA}{\mbox{\sc MA}}
\newcommand{\GMI}{\mbox{\sc GMI}}
\newcommand{\cGMI}{\mbox{\sc Coloured-GMI}}
\newcommand{\GMA}{\mbox{\sc GMA}}
\newcommand{\LMA}{\mbox{\sc LMA}}
\newcommand{\LMI}{\mbox{\sc LMI}}
\newcommand{\cLMI}{\mbox{\sc Coloured-LMI}}

\newcommand{\mi}{\mbox{\sc Matroid Isomorphism}}
\newcommand{\lmi}{\mbox{\sc Linear Matroid Isomorphism}}
\newcommand{\gmi}{\mbox{\sc Graphic Matroid Isomorphism}}
\newcommand{\pmi}{\mbox{\sc Planar Matroid Isomorphism}}

\newcommand{\ma}{\mbox{\sc Matroid Automorphism}}

\newenvironment{aclaim}[1]{\noindent {\bf Claim~#1:~}\em }{\smallskip}

\begin{document}
\title{\sf On the Complexity of Matroid Isomorphism Problem}
\author{Raghavendra Rao B.V.\thanks{ 
        {\it The Institute of Mathematical Sciences,} 
        {\it C.I.T. Campus,} 
        {\it Chennai 600 113, India. }
        {\tt bvrr@imsc.res.in} }
       \and
       Jayalal Sarma M.N.
       \thanks{ 
       {\it Institute for Theoretical Computer Science,} 
        {\it Tsinghua University,} 
        {\it Beijing 100 084, China. }
        {\tt jayalal@tsinghua.edu.cn}  
}\thanks{
The work was done when this author was also a
         graduate student at the Institute of Mathematical Sciences, 
         Chennai, India.}
}

\date{} 
\maketitle
\begin{abstract}
 
  We study the complexity of testing if two given matroids are
  isomorphic. The problem is easily seen to be in $\Sigma_2^p$. In the
  case of linear matroids, which are represented over polynomially
  growing fields, we note that the problem is unlikely to be
  $\Sigma_2^p$-complete and is $\co\NP$-hard. We show that when the
  rank of the matroid is bounded by a constant, linear matroid
  isomorphism, and matroid isomorphism are both polynomial time
  many-one equivalent to graph isomorphism.
  
  We give a polynomial time Turing reduction from graphic matroid
  isomorphism problem to the graph isomorphism problem. Using this, we
  are able to show that graphic matroid isomorphism testing for planar
  graphs can be done in deterministic polynomial time. We then give a
  polynomial time many-one reduction from bounded rank matroid
  isomorphism problem to graphic matroid isomorphism, thus showing
  that all the above problems are polynomial time equivalent.

  Further, for linear and graphic matroids, we prove that the
  automorphism problem is polynomial time equivalent to the
  corresponding isomorphism problems. In addition, we give a
  polynomial time membership test algorithm for the automorphism group
  of a graphic matroid.
\end{abstract}


\section{Introduction}

Isomorphism problems over various mathematical structures have been a
source of intriguing problems in complexity theory (see~\cite{AT05}).
The most important problem of this domain is the well-known graph
isomorphism problem. Though the complexity characterization of the
general version of this problem is still unknown, there have been
various interesting special cases of the problem which are known to
have polynomial time algorithms \cite{BGM82,Luk80}. In this paper we 
talk about isomorphism problem associated with matroids.

A matroid $M$ is a combinatorial object defined over a finite set $S$
(of size $m$) called the {\em ground set}, equipped with a non-empty
family $\I$ of subsets of $S$ (containing the empty subset) which is
closed under taking of subsets and satisfies the {\em exchange axiom}
: for any $I_1, I_2 \in \I$ such that $|I_1| > |I_2|$, $\exists x \in
I_1 \setminus I_2$, $I_2 \cup \{x\} \in \I$. The sets in $\I$ are
called {\em independent sets}. The rank of the matroid is the size of
the maximal independent set.
This provides useful abstractions of many concepts in combinatorics
and linear algebra~\cite{Whi35}. The theory of matroids is a well
studied area of combinatorics~\cite{Oxley-Book}. We study the problem 
of testing isomorphism between two given matroids. 

Two matroids $M_1$ and $M_2$ are said to be isomorphic if there is a
bijection between the elements of the ground set which maps
independent sets to independent sets, (or equivalently circuits to
circuits, or bases to bases, see section~\ref{sec:prelim}). Quite
naturally, the representation of the input matroids is important in
deciding the complexity of the algorithmic problem.

There are several equivalent representations of a matroid. For
example, enumerating the maximal independent sets (called bases) or
the minimal dependent sets (called circuits) also defines the matroid.
These representations, although can be exponential in the size of the
ground set, indeed exist for every matroid, by definition. With this
enumerative representation, Mayhew~\cite{May08} studied the matroid
isomorphism problem, and shows that the problem is equivalent to graph
isomorphism problem. However, a natural question is whether the
problem is difficult when the representation of the matroid is
more implicit?. In a black-box setting, one can also consider the input
representation in the form of an oracle or a black-box, where the
oracle answers whether a given set is independent or not.

More implicit (and efficient) representation of matroids have been
studied. One natural way is to identify the given matroid 
with matroids defined over combinatorial or algebraic
objects which have implicit descriptions. A general framework in this
direction is the representation of a matroid over a field. A matroid $M
= (S,\mathcal{I})$ of rank $r$ is said to be {\em representable} over a
field $\mathbb{F}$ if there is a 
  map, $\phi : S \to
\mathbb{F}^r$ such that, $\forall A \subseteq S$, $A \in \mathcal{I}
\iff \phi(A)$ is linearly independent over $\mathbb{F}^r$ as a vector
space. However, there are matroids which do not admit linear
representations over any field. (For example, the  Vam\'os Matroid, See
Proposition 6.1.10,~\cite{Oxley-Book}.). In contrast, there are
matroids (called regular matroids) which admit linear representations
over all fields.

Another natural representation for a matroid is over graphs. For any
graph $X$, we can associate a matroid $M(X)$ as follows: the set of
edges of $X$ is the ground set, and the acyclic subgraphs of the given
graph form the independent sets. A matroid $M$ is called a {\em graphic
matroid} (also called polygon matroid or cyclic matroid) if it is
isomorphic to $M(X)$ for some graph $X$. It is known that graphic
matroids are linear. Indeed, the incidence matrix of the graph will
give a representation over $\F_2$. There are linear matroids which are
not graphic. (See~\cite{Oxley-Book} for more details.)

The above definitions themselves highlight the importance of testing
isomorphism between two given matroids.  
We study the isomorphism problem for the case of linear matroids (Linear
Matroid Isomorphism problem ($\LMI)$ and graphic matroids (Graphic
Matroid Isomorphism problem ($\GMI$)).

From a complexity perspective, the general case of the problem is in
$\Sigma_2^p$. However, it is not even clear a priori if the problem is
in $\NP$ even in the above restricted cases where there are implicit
representations. But we note that for the case of graphic matroids the
problem admits an $\NP$ algorithm. Hence an intriguing question is
about the comparison of this problem to the well studied graph
isomorphism problem.

At an intuitive level, in the graph isomorphism problem we ask for a
map between the vertices that preserves the adjacency relations,
whereas in the case of graphic matroid isomorphism, we ask for maps
between the edges such that the set of cycles (or spanning trees) in
the graph are preserved. As an example, in the case of trees, any
permutation gives a 2-isomorphism, where as computing the isomorphism
between trees is known to be $\L$-complete. This indicates that the
reduction between the problems cannot be obtained by a local
replacement of edges with gadgets, and has to consider the global
structure.

An important result in this direction, due to Whitney
(see~\cite{Whi32}), says that in the case of 3-connected graphs, the
graphs are isomorphic if and only if the corresponding matroids are
isomorphic (see section~\ref{sec:graphs}). Thus the problem of testing
isomorphism of graphs and the corresponding graphic matroids are
equivalent for the case of 3-connected graphs are equivalent. Despite
this similarity between the problems, to the best of our knowledge,
there has not been a systematic study of $\GMI$ and its relationships
to graph isomorphism problem ($\GI$). This immediately gives a
motivation to study the isomorphism problem for 3-connected graphs.
In particular, from the recent results on graph isomorphism problem
for these classes of graphs \cite{DLN08,TW08}, it follows that
graphic matroid isomorphism problem for 3-connected planar graphs
$\L$-complete.

In this context we study the general, linear and graphic matroid
isomorphism problems. Our main contributions in the paper are as
follows:

\begin{itemize}
\item Matroid isomorphism problem is easily seen to be in
  $\Sigma_2^p$.  In the case of linear matroids where the field is
  also a part of the input we observe that the problem is
  $\co\NP$-hard (Proposition~\ref{prop:conphard}), and is unlikely to
  be $\Sigma_2^p$-complete (Proposition~\ref{prop:MIhard}). We also
  observe that when the rank of the matroid is bounded, linear matroid
  isomorphism, and matroid isomorphism are both equivalent to $\GI$
  (Theorem~\ref{thm:lmib-gi})\footnote{We note that, although not
    explicitly stated, the equivalence of bounded rank matroid
    isomorphism and and graph isomorphism also follows from the
    results of Mayhew~\cite{May08}. However, it is not immediately
    clear if the GI-hard instances are linearly representable. 
    Our proofs are different and extends this to linear matroids.}
\item We develop tools to handle colouring of ground set elements in
  the context of isomorphism problem. We show that coloured version of
  the linear matroid isomorphism and graphic matroid isomorphisms are
  as hard as the general version
  (Lemma~\ref{lem:lmi-col},~\ref{lem:coloring}). As an immediate
  application of this, we show that the automorphism problems for
  graphic matroids and linear matroids are polynomial time Turing
  equivalent to the corresponding isomorphism problems. In this
  context, we also give a polynomial time membership test algorithm
  for the automorphism group of a graphic matroid
  (Theorem~\ref{thm:iso-auto}).

\item We give a polynomial time Turing reduction from graphic matroid
  isomorphism problem to the graph isomorphism problem by developing
  an edge colouring scheme which algorithmically uses a decomposition
  given by \cite{HT73} (and \cite{CE80}) and reduce the graphic
  matroid 
  (Theorem~\ref{thm:gmi-to-gi}).  Our reduction, in particular implies
  that the graphic matroid isomorphism testing for planar graphs can
  be done in deterministic polynomial time
  (Corollary~\ref{cor:planarMI}).


\item Finally, we give a reduction from bounded rank matroid
  isomorphism problem to graphic matroid isomorphism
  (Theorem~\ref{thm:bmi-gmi}), thus showing that all the above
  problems are poly-time equivalent. 
\end{itemize}
Table~\ref{tab:comp-MI} below summarizes the complexity of matroid
  isomorphism problem under various input representations. 
\begin{table}[h]
\begin{center}
\begin{tabular}{|l|l|c|}
\hline
Repn. of $M_1, M_2$ & Complexity Bounds for $\MI$ \\
\hline
List of Ind. sets & $\GI$-complete \cite{May08}\\
\hline
Linear & $\GI$-hard, $\co\NP$-hard (\cite{Hli07,OW02}).  \\
\hline
Linear (bounded rank) & $\GI$ complete \\
\hline
Graphic & Turing equivalent to $\GI$ \\
\hline
Planar & $\P$ \\
\hline
Planar 3-connected  & $\L$-complete \\
\hline

\end{tabular}
\vspace{1mm}
\caption{Complexity of $\MI$ under various input representations}
\label{tab:comp-MI}
\end{center}
\end{table}

\section{Notations and Preliminaries}
\label{sec:prelim}

All the complexity classes used here are standard and we refer the
reader to any standard text book (for e.g. see~\cite{Gol08}).  Now we
collect some basic definitions on matroids (see also
\cite{Oxley-Book}). Formally, a matroid $M$ is a pair $(S,\I)$, where
$S$ is a finite set called the ground set of size $m$ and $\I$ is a
collection of subsets of $S$ such that: (1) the empty set $\phi$, is
in $\I$.  (2) If $I_1 \in I$ and $I_2 \subset I_1$, then $I_2 \in
\I$. (3) If $I_1,I_2 \in \I$ with $|I_1| < |I_2|$, then $\exists x \in
I_2 \setminus I_1$ such that $I_1 \cup \{x\}$ is in $\I$.

The $\rank$ function of a matroid is a map {\sf rank}:~$2^{S} \to
\mathbb{N}$, is defined for a $T \subseteq S$, as the maximum size of
any element of $\I$ that is contained in $T$. The rank of the matroid
is the maximum value of this function. A {\em circuit} is a minimal
dependent set.  Spanning sets are subsets of $S$ which contains at
least one basis as its subset. Notice that a set $X \subseteq S$ is
spanning if and only if $rank(X) = rank(S)$. Moreover, $X$ is a basis
set if and only if it is a minimal spanning set.  For any $F \subseteq
S$, $cl(F) = \{x \in S~:~rank(F \cup x) = ~rank(F) \}$.  A set $F
\subseteq S$ is a flat if $cl(F) = F$. {\em Hyperplanes} are flats
which are of rank $r-1$, where $r=\rank(S)$.
  $X \subseteq S$ is a hyperplane if and only
if it is a maximal non-spanning set.

An isomorphism between two matroids $M_1$ and $M_2$ is a bijection
$\phi : S_1 \to S_2$ such that $\forall C \subseteq S_1: C \in \C_1~
\iff ~\phi(C) \in \C_2$, where $\C_1$ and $\C_2$ are the family of
circuits of the matroids $M_1$ and $M_2$ respectively. Now we state
the computational problems more precisely.

\begin{problem}[$\mi(\MI)$]
Given two matroids $M_1$ and $M_2$ as their independent set oracles,
does there exist an isomorphism between the two matroids?
\end{problem}

Given a matrix $A$ over a field $\F$, we can define a matroid $M[A]$
with columns of $A$ as the ground set and linearly independent columns
as the independent sets of $M[A]$. A matroid $M=(E,\I)$ with rank$=r$
is said to be representable over $\F$, if there is a
 map
$\Phi:E\to\F^r$ such that $I\in \I \iff \Phi(I)$ is linearly
independent in $\F^n$.  {\it Linear matroids} are matroids
representable over fields. Without loss of generality we can assume
that the representation is of the form of a matrix where the columns
of the matrix correspond to the ground set elements. We assume that
the field on which the matroid is represented is also a part of the
input, also that the field has at least $m$ elements and at most
$poly(m)$ elements, where $m=poly(n)$.

\begin{problem}[$\lmi(\LMI)$]
Given two matrices $A$ and $B$ over a given field $\F$ does there
exist an isomorphism between the two linear matroids represented by
them?.
\end{problem}

As mentioned in the introduction, given a graph $X = (V,E)$ ($|V|=n,
|E|=m$), a classical way to associate a matroid $M(X)$ with $X$ is to
treat $E$ as ground set elements, the bases of $M(X)$ are spanning
forests of $X$. Equivalently circuits of $M(X)$ are simple cycles in
$X$. A matroid $M$ is called {\em graphic} iff~ $\exists X$ such that
$M=M(X)$.

Evidently, adding vertices to a graph $G$ with no incident edges will
not alter the matroid of the graph.  Without loss of generality
we can assume that $G$ does not have self-loops.


\begin{problem}[$\gmi(\GMI)$]
Given two graphs $X_1$ and $X_2$ does there exist an isomorphism
between $M(X_1)$ and $M(X_2)$?.
\end{problem}

Another associated terminology in the literature is about
2-ismorphism.  Two graphs $X_1$ and $X_2$ are said to {\em
  $2$-isomorphic} (denoted by $X_1 \cong_2 X_2$) if their
corresponding graphic matroids are isomorphic. Thus the above problem
asks to test if two given graphs are 2-isomorphic.

In a rather surprising result, Whitney~\cite{Whi33} came up with a
combinatorial characterisation of 2-isomorpic graphs. We briefly
describe it here. Whitney defined the following operations.

\begin{itemize}
\item {\em Vertex Identification:} Let $v$ and $v'$ be vertices of
  distinct components of $X$. We modify $X$ by identifying $v$ and $v'$
  as a new vertex $\bar{v}$.
\item {\em Vertex Cleaving:} This is the reverse operation of vertex
  identification so that a graph can only be cleft at the a cut-vertex
  or at a vertex incident with a loop.
\item {\em Twisting:} Suppose that the graph $X$ is obtained from the
  disjoint graphs $X_1$ and $X_2$ by identifying vertices $u_1$ of
  $X_1$ and $u_2$ of $X_2$ as the vertex $u$ of $X$, identifying
  vertices $v_1$ of $X_1$ and $v_2$ of $X_2$ as the vertex $v$ of $X$.
  In a {\em twisting}  of $X$ about $\{u,v\}$, we identify,
  instead $u_1$ with $v_2$ and $u_2$ with $v_1$ to get a new graph $X'$.
\end{itemize}

\begin{theorem}[Whitney's 2-ismorphism theorem](\cite{Whi33}, see also \cite{Oxley-Book})
  \label{thm:Whi33}
  Let $X_1$ and $X_2$ be two graphs having no isolated vertices. Then
  $M(X_1)$ and $M(X_2)$ are isomorphic if and only if $X_1$ can be
  transformed to a graph isomorphic to $X_2$ by a sequence of
  operations of vertex identification, cleaving and/or twisting.
\end{theorem}

The graphic matroids of planar graphs are called {\em planar matroids}.
We now define the corresponding isomorphism problem for graphic
matroids,

\begin{problem}[$\pmi(\PMI)$]
Given two planar graphs $X_1$ and $X_2$ does there exist an
isomorphism between their graphic matroids ?.
\end{problem}

As a basic complexity bound, it is easy to see that $\MI \in
\Sigma_2^p$. Indeed, the algorithm will existentially guess a
bijection $\sigma:S_1 \to S_2$ and universally verify if for every
subset $C \subseteq S_1$, $C \in \C_1 \iff \sigma(C) \in \C_2$ using
the independent set oracle.

%
\section{Linear Matroid Isomorphism} 

In this section we present some observations and results on
$\lmi$. Some of these follow easily from the techniques in the
literature. We make them explicit in a form that is relevant to the
problem that we are considering.

We first observe that using the arguments similar to that of
\cite{KST93} one can show $\overline{\LMI}\in \BP.\Sigma_2^\P$ (Notice
that an obvious upper bound for this problem is $\Pi_2$). We include
some details of this here while we observe some points about the
proof.
\begin{proposition}
\label{lem:MI-Sigma2}
$\overline{\LMI}\in \BP.\Sigma_2^\P$
\end{proposition}

\begin{proof}
  Let $M_1$ and $M_2$ be the given linear matroids having $m$ columns
  each.  We proceed as in \cite{KST93}, for the case of $\GI$. To give
  a $\BP.\Sigma_2^\P$ algorithm for $\overline{\LMI}$, define the
  following set:

  \[ N(M_1,M_2) = \left\{ (N,\phi) : (N \cong M_1) \lor (N \cong M_2)
    \land \phi \in Aut(N) \right\} \]

  where $Aut(H)$ contains all the permutations (bijections) which are
  isomorphisms of matroid $N$ to itself. The key property that is used
  in \cite{KST93} has the following easy counterpart in our context.

  For any matroid $M$ on a ground set of size $m$, if $Aut(M)$ denote
  the automorphism group of $M$, $\# M$ denotes the number of
  different matroids isomorphic to $M$, $|Aut(M)| * (\#M) = |S_m|$.

  \[ M_1 \cong M_2 \implies |N(M_1,M_2)| = m! \]
  \[ M_1 \not\cong M_2 \implies |N(M_1,M_2)| = 2.m! \]

  As in \cite{KST93}, we can amplify this gap and then using a good
  hash family and utilise the gap to distinguish between the two
  cases.  In the final protocol (before amplifying) the verifier
  chooses a hash function and sends it to the prover, the prover returns
  a tuple $(N,\phi)$ along with a proof that this belongs to
  $N(M_1,M_2)$. (Notice that this will not work over very large
  fields, especially over infinite fields.) Verifier checks this
  claim along with the hash value of the tuple. This can be done in
  $\Sigma_2^p$. Hence the entire algorithm gives an upper bound of
  $\BP.\exists.\Sigma_2^p = \BP.\Sigma_2^p$, and thus the result
  follows.
\end{proof}

Now, we know that~\cite{Sch89}, if $\Pi_2^p \subseteq
\BP.\Sigma_2^p$ then $\PH = \BP.\Sigma_2^p = \Sigma_3^p$. 
Thus we get the following:

\begin{theorem}
  \label{prop:MIhard} 
  $\LMI \in \Sigma_2^p$. ~In addition, $\LMI \textrm{ \em is }
  \Sigma_2^\P\textrm{\em -hard} \implies \PH = \Sigma_3^\P$.
\end{theorem}

We notice that a special case of this is already known to be
$\co\NP$-hard. A matroid of rank $k$ is said to be {\em uniform} if
all subsets of size at most $k$ are independent.  Testing if a given
linear matroid of rank $k$ is uniform is known to be
$\coNP$-complete~\cite{OW02}.  We denote them by $U_{k,m}$, the
uniform matroid whose ground set is of $m$ elements. Now notice that
the above result is equivalent to checking if the given linear matroid
of rank $k$ is isomorphic to $U_{k,m}$. To complete the argument, we
use a folklore result that $U_{k,m}$ is representable over any field
$\F$ which has at least $m$ non-zero elements. We give some details
here since we have not seen an explicit description of this in the
literature.

\begin{claim}
  \label{prop:repn-uniform}
  Let $|\F| > m$, $U_{k,m}$ has a representation over $\F$.
\end{claim}
\begin{proof}
  Let $\{ \alpha_1, \ldots, \alpha_m \}$ be distinct elements of $\F$,
  and $\{ s_1, \ldots, s_m \}$ be elements of the ground set of
  $U_{k,m}$ Assign the vector $(1, \alpha_i, \alpha_i^2, \ldots,
  \alpha_i^{k-1}) \in \F^k$ to the element $s_i$. Any $k$ subset of
  these vectors forms a Vandermonde matrix, and hence linearly
  independent. Any larger set is dependent since the vectors are in
  $\F^k$.
\end{proof}

This gives us the following proposition.
\begin{proposition}
\label{prop:conphard}
$\LMI$ is $\co\NP$-hard.
\end{proposition}

The above proposition also holds when the representation is over
infinite fields. In this case, the proposition also more directly
follows from a result of Hlinen\'y \cite{Hli07}, where it is shown
that the problem of testing if a spike (a special kind of matroids)
represented by a matrix over $\Q$ is the free spike is $\co\NP$
complete. He also derives a linear representation for spikes.

Now we look at bounded rank variant of the problem. We denote by
$\LMI_b$ ($\MI_b$), the restriction of $\LMI$ ($\MI$) for which the
input matrices have rank bounded by $b$. In the following we use the
following construction due to Babai \cite{Bab78} to prove $\LMI_b
\equiv_m^p \GI$.

Given a graph $X=(V,E)$ ($3\leq k \leq d $, where, $d$ is the minimum vertex
degree of $X$), define a matroid $M=St_k(X)$ of rank $k$ with the
ground set as $E$ as follows: every subset of $k-1$ edges is
independent in $M$ and every subset of $E$ with $k$ edges is
independent if and only if they do not share a common vertex. Babai
proved that $Aut(X)\cong Aut(St_k(X))$ and also gave a linear
representation for $St_k(X)$ (Lemma 2.1 in \cite{Bab78}) for all $k$
in the above range.

\begin{theorem}
  \label{thm:lmib-gi}
  For any constant $b \ge 3$, $\LMI_b \equiv_m^p \GI$.
\end{theorem}
\begin{proof}
  {\sc $\GI \le_m^p \LMI_b$:}  Let $X_1=(V_1,E_1)$ and
  $X_2=(V_2,E_2)$ be the given $\GI$ instance. We can assume that the
  minimum degree of the graph is at least $3$ since otherwise we can
  attach cliques of size $n+1$ at every vertex. We note that from
  Babai's proof we can derive the following stronger conclusion.
  \begin{lemma} 
   \label{lem:stk}
    $X_1 \cong X_2 \iff \forall~k \in [3,d],~St_k(X_1)\cong St_k(X_2)$
  \end{lemma}
  \begin{proof}
    Suppose $X_1 \cong X_2$ via a bijection $\pi: V_1 \to V_2$. (The
    following proof works for any $k \in [3,d]$.) Let $\sigma: E_1 \to
    E_2$ be the map induced by $\pi$. That is $\sigma(\{u,v\}) =
    \{\pi(u),\pi(v)\}$. Consider an independent set $I \subseteq E_1$
    in $St_k(X_1)$. If $|I| \le k-1$ then $|\sigma(I)| \le k-1$ and
    hence $\sigma(I)$ is independent in $St_k(X_2)$. If $|I| = k$, and
    let $\sigma(I)$ be dependent. This means that the edges in
    $\sigma(I)$ share a common vertex $w$ in $X_2$. Since $\pi$ is an
    isomorphism which induces $\sigma$, $\pi^{-1}(w)$ must be shared
    by all edges in $I$. Thus $I$ is independent if and only if
    $\sigma(I)$ is independent.  Suppose $St_k(X_1) \cong St_k(X_2)$
    via a bijection $\sigma: E_1 \to E_2$. By definition, any subset
    $H \subseteq E_1$ is a hyperplane of $St_k(X_1)$ if and only if
    $\sigma(H)$ is a hyperplane of $St_k(X_2)$. Now we use the
    following claim which follows from \cite{Bab78}. 

  \begin{claim}[\cite{Bab78}]
    \label{claim:babai}
    For any graph $X$, any dependent hyperplane in $St_k(X)$ 
    is a maximal set of edges which share a common vertex (forms a
    star) in $X$, and these are the only dependent hyperplanes.
  \end{claim}

  Now we define the graph isomorphism $\pi : V_1 \to V_2$ as follows.
  For any vertex $v$, look at the star $E_1(v)$ rooted at $v$, we know
  that $\sigma (E_1(v)) = E_2(v')$ for some $v'$. Now set $\pi(v) =
  v'$. From the above claim, $\pi$ is an isomorphism.
  \end{proof}

  It remains to show that representation for $St_k(X)$ ($X=(V,E)$) can
  be computed in polynomial time. We choose $k=3$ (by the above proof,
  $\exists k$ and $\forall k$ in the Lemma \ref{lem:stk} are
  equivalent). Now we show that the representation of $St_k(X)$ given
  in \cite{Bab78} is computable in polynomial time.  The
  representation of $St_k(X)$ is over a field $\F$ such that
  $|\F|\ge|V|^{2k-1}$. For $e=\{u,v\}\in E$ assign a vector $b_e=[1,
    (x_u+x_v), (x_ux_v), y_{e,1}, \ldots, y_{e,k-3}] \in \F^k$, where
  $x_u,x_v$ and $y_{e,i}$ are distinct unknowns. To represent
  $St_k(X)$ we need to ensure that the $k$-subsets of the columns
  corresponding to a basis form a linearly independent set, and all
  the remaining $k$-subsets form a dependent set. Babai \cite{Bab78}
  showed that by the above careful choice of $b_e$, it will be
  sufficient to ensure only the independence condition. He also proved
  the existence of a choice of values for the variables which achieves
  this if $|\F|\ge|V|^{2k-1}$.

  We make this constructive. As $k$ is a constant, the number of
  bases is bounded by $\poly(m)$.  We can greedily
  choose the value for each variable at every step, such that on
  assigning this value, the resulting set of constant ($k\times k$)
  size matrices are non-singular. Since there exists a solution, this
  algorithm will always find one. Thus we can compute a representation
  for $St_k(X)$ in polynomial time.

  {\sc $\LMI_b \le_m^p \GI$:} Let $A_{k \times m}$ and $B_{k \times
    m}$ be two matrices of rank $b$ at the input. Now define the
    following bipartite graph $X_A = (U_A, V_A, E_A)$ (similarly for
    $X_B$), where $U_A$ has a vertex for each column of $A$, and $V_A$
    has a vertex for each maximal independent set of $A$ (Notice that
    there are at most ${m \choose b}=O(m^b)$ of them) and $\forall
    i\in U_A, I\in V_A,\ \{i,I\}\in E_A \iff i\in I$. Now we claim that
    $M(A) \cong M(B) \iff$ $X_A \cong X_B$ where the isomorphism maps
    $V_A$ to $V_B$, and which is reducible to $\GI$. It is easy to see
    that the matroid isomorphism can be recovered from the map between
    the sets.
\end{proof}

Observe that the reduction $\LMI_b \le_m^p \GI$ can be done even if
the input representation is an independent set oracle. This gives the
following corollary.
\begin{corollary}
$\LMI_b \equiv_m^p \MI_b \equiv_m^p \GI$.
\end{corollary}

\section{Isomorphism Problem of Coloured Matroids}
\label{sec:coloring}

Vertex or edge colouring is a classical tool used extensively in
proving various results in graph isomorphism problem. We develop
similar techniques for matroid isomorphism problems too.

An edge-$k$-colouring of a graph $X=(V,E)$ is a function $f: E \to \{1,
\ldots, k \}$.  Given two coloured graphs
$X_1=(V_1,E_1,f_1)$ and $X_2 = (V_2,E_2,f_2)$, the $\cGMI$ asks for an
isomorphism which preserves the colours of the edges.
%
Not surprisingly,  
we can prove the following.

\begin{lemma}
\label{lem:coloring}
$\cGMI$ is $\AC^0$ many-one reducible to $\GMI$.
\end{lemma}
\begin{proof}
  Let $X_1=(V_1,E_1,f_1)$ and $X_2 = (V_2,E_2,f_2)$, be the two
  $k$-coloured graphs at the input, with $n=|V_1|=|V_2|$. For every
  edge $e=(u,v) \in E_1$ (respectively $E_2$), add a path
  $P_e=\{(u,v_{e,1}),(v_{e,1},v_{e,2}),\ldots, (v_{e,n+f_1(e)},v)\}$ of length
  $n+f_1(e)$ (respectively $n+f_2(e)$)
  Where $v_{e,1}, \ldots v_{e,n+f_1(e)}$ are new vertices.  Let $X_1'$
  and $X_2'$ be the two new graphs thus obtained. By definition, any
  2-isomorphism between $X_1'$ and $X_2'$ can only map cycles of equal
  length to themselves. There are no simple cycles of length more than
  $n$ in the original graphs. Thus, given any 2-isomorphism between
  $X_1'$ and $X_2'$, we can recover a 2-isomorphism between $X_1$ and
  $X_2$ which preserves the colouring and vice versa.
\end{proof}

Now we generalize the above construction to the case of linear matroid
isomorphism. $\cLMI$ denotes the variant of $\LMI$ where the inputs
are the linear matroids $M_1$ and $M_2$ along with  colour functions
$c_i:\{1,\ldots,m\} \to \N, i\in\{1,2\} $. The problem is to test
if there is an isomorphism between $M_1$ and $M_2$ which preserves the
colours of the column indices. We have,

\begin{lemma} 
\label{lem:lmi-col}
$\cLMI$ is $\AC^0$ many-one reducible to $\LMI$.
\end{lemma}

\begin{proof}
  Let $M_1$ and $M_2$ be two coloured linear matroids represented over
  a field $\F$. We illustrate the reduction where only one column
  index of $M_1$ (resp. $M_2$) is coloured. Without loss of generality,
  we assume that there are no two vectors in $M_1$ (resp.$M_2$) which
  are scalar multiples of each other.

  We transform $M_1$ and $M_2$ to get two matroids $M_1'$ and
  $M_2'$. In the transformation, we add more columns to the matrix
  (vectors to the ground set) and create dependency relations in such
  a way that any isomorphism between the matroids must map these new
  vectors in $M_1$ to the corresponding ones $M_2$.

  We describe this transformation in a generic way for a matroid
  $M$. Let $\{ e_1,\ldots,e_m \}$ be the column vectors of $M$, where
  $e_i \in \F^n$. Let $e=e_1$ be the coloured vector in $M$.

  Choose $m' > m$, we construct $\ell=m+m'$ vectors $f_1, \ldots
  f_\ell \in \F^{n+m'}$ as the columns of the following $(n+m') \times
  \ell$ matrix. The $i^\mathrm{th}$ column of the matrix represents
  $f_i$.
   \[
   \left[
   \begin{array}{cccc|cccccccc}
    e_{11} & e_{21} & \ldots & e_{m1} & e_{11} & 0 & \ldots & 0 & 0 & \ldots & 0 \\
    e_{12} & e_{22} & \ldots & e_{m2} & 0  & e_{12} & \ldots & 0 & 0 & \ldots & 0 \\
    \vdots & \vdots &\ddots & \vdots  & \vdots & \vdots & \ddots & \vdots & \vdots  
    & \ddots & \vdots \\
    e_{1m} & e_{2m} & \ldots & e_{mm} & 0 & 0 & \ldots & e_{1m} &  0 & \ldots & 0  \\
    \hline
    0 & 0 & \ldots & 0 & 1 & -1 & 0 & 0 & \hdotsfor{2}& 0 \\
    \vdots & \vdots  & \ldots & \vdots & 0 & 1 & -1 & 0 & \hdotsfor{2} & 0  \\
    \vdots & \vdots &\ldots & \vdots  & \vdots & \vdots & \ddots & \ddots & \hdotsfor{2}
    & \vdots \\
    0 & 0 & \ldots & 0 & 0 & 0 & \ldots && 0 & 1 & -1\\
    0 & 0 & \ldots & 0 & -1 & 0 & \ldots && 0 & 0 & 1\\
   \end{array}
   \right]
   \]
  where $-1$ denotes the additive inverse of $1$ in $\F$. Denote the
  above matrix as $M' = \begin{pmatrix} A & B \\ C & D \end{pmatrix}$.
  Let $S = \{ f_{m+1},\ldots,f_{m+m'} \}$. We observe the following:
  \begin{enumerate}
  \item \label{obs:1}
    Columns of $B$ generate $e_1$. Since $C$ is a $0$-matrix $f_1
    \in Span(S)$.
  \item \label{obs:2}Columns of $D$ are minimal dependent. Any proper subset of
      columns of $D$ will split the $1$, $-1$ pair in at least a row
      and hence will be independent.
    \item \label{obs:3} $S$ is linearly independent. Suppose not. Let
      $\sum_{i=m}^{m+m'} \alpha_i f_i = 0$. Restricting this to the
      columns of $B$ gives that $\alpha_j = 0$ for first $j$ such that
      $e_{1j} \ne 0$.  Thus this gives a linearly dependent proper
      subset of columns of $B$, and contradicts the above observation.
  \item \label{obs:4}If for any $f \notin S$, $f = \sum_{f_i \in S} \alpha_i f_i$, 
      then $\alpha_i$'s must be the same.
  \end{enumerate}

  Now we claim that the newly added columns respect the circuit
  structure involving $e_1$. Let $\mathcal{C}$ and $\mathcal{C}'$
  denote the set of circuits of $M$ and $M'$ respectively.\\

  \begin{claim}
    \label{claim:lmi-col}
    \begin{eqnarray*}
      \left\{ e_1, e_{i_2}, \ldots, e_{i_k} \right\}  
      \in \mathcal{C} & \iff & \left\{ f_1, f_{i_2}\ldots, f_{i_k} \right\} \in \mathcal{C}' 
      \textrm{ and }\\
      & & \left\{f_{i_2}, \ldots, f_{i_k}, f_{m+1},\ldots, f_{m+m'}\right\} \in \mathcal{C}'
    \end{eqnarray*}
  \end{claim}
  \begin{proof}
    Suppose $c = \{ e_1, e_{i_2}, \ldots, e_{i_k} \}$ is a circuit in
    $M$. Then clearly $\{f_1, f_{i_2}, \ldots, f_{i_k}\}$ is a
    cycle, since they are nothing but vectors in $c$ extended with
    $0$s.  Since $\{ f_{i_2},\ldots, f_{i_k} \} $ and
    $\{f_{m+1},\ldots, f_{m+m'} \}$ both generate $f_1$, the set $F =
    \{f_{i_2},\ldots, f_{i_k},f_{m+1},\ldots, f_{m+m'}\}$ is a
    linearly dependent set. Now we argue that $F$ is a minimal
    dependent set, and hence is a circuit. Denote by $G$ the set
    $\{f_{i_2},\ldots, f_{i_k}\}$.

    Suppose not, let $F' \subset F$ be linearly dependent. Since $S$
    is linearly independent (property~\ref{obs:3} above), we note that
    $F' \not\subseteq \{f_{m+1},\ldots, f_{m+m'}\}$.
    Therefore, $f_{i_j} \in F'$ for some $0 \le j \le k$. Since $F'$
    is dependent, express $f_j$ in terms of the other elements in
    $F'$:
    \[ f_j = \sum_{g \in G} \gamma_g g +\sum_{s\in S}\delta_s s\]
    Since $G$ is linearly independent, at least one of the $\delta_s$
    should be non-zero.
    Restrict this to the matrices $C$ and $D$. This
    gives a non-trivial dependent proper subset of $D$ and hence
    a contradiction.
  \end{proof}

  From the above two observations and the fact that there is no other
  column in $M$ which is a multiple of $e$, the set $f(e) =
  \{f_1,f_{m+1},\ldots, f_{m+m'}\}$ is a unique circuit of length
  $m'+1$ in $M'$, where $e$ is column which is coloured.

  Now we argue about the isomorphism between $M_1'$ and $M_2'$
  obtained from the above operation, and there is a unique circuit of
  length $m'+1 > m$ in both $M_1'$ and $M_2'$ corresponding to two
  vectors $e \in M_1$ and $e' \in M_2$. Hence any matroid isomorphism
  should map  these sets to each other. From such an isomorphism,
  we can recover the a matroid isomorphism between $M_1$ and $M_2$
  that maps between $e$ and $e'$, thus preserving the colours. Indeed,
  if there is a matroid isomorphism between $M_1$ and $M_2$, that can
  easily be extended to $M_1'$ and $M_2'$. 
  
  For the general case, let $k$ be the number of different colour
  classes and $c_i$ denote the size of the $i$th colour class. Then for
  each vector $e$ in the color class $i$, we add $l_i=m+m'+i$ many new
  vectors, which also increases the dimension of the space by
  $l_i$. Thus the total number of vectors in the new matroid is
  $\sum c_i(l_i)\le m^3$. Similarly, the dimension of the space 
  is bounded by $m^3$. This completes the proof of
  Lemma~\ref{lem:lmi-col}.
  
\end{proof}

\section{Graphic Matroid Isomorphism}
\label{sec:graphs}

In this section we study $\GMI$.  Unlike in the case of the graph
isomorphism problem, an $\NP$ upper bound is not so obvious for
$\GMI$. We start with the discussion of an $\NP$ upper bound for
$\GMI$. 

As stated in Theorem~\ref{thm:Whi33}, Whitney gave an exact
characterization of when two graphs are 2-isomorphic, in terms of
three operations; twisting, cleaving and identification. Note that it
is sufficient to find 2-isomorphisms between 2-connected components of
$X_1$ and $X_2$. In fact, any {\em matching} between the sets of
2-connected components whose edges connect 2-isomorphic components
will serve the purpose. This is because, any 2-isomorphism preserves
simple cycles, and any simple cycle of a graph is always within a
2-connected component. Hence we can assume that both the input graphs
are 2-connected and in the case of 2-connected graphs, twist is the
only possible operation.

The set of separating pairs does not change under a twist operation.
Despite the fact that the twist operations need not commute, Truemper
\cite{Tru80} gave the following bound.

\begin{lemma}[\cite{Tru80}]
Let $X$ be a 2-connected graph of $n$ vertices, and let $Y$ be a graph
$2$-isomorphic to $X$, then: $X$ can be transformed to graph $X'$
isomorphic to $Y$ through a sequence  at most $n-2$
twists.
\end{lemma}

Using this lemma we get  an $\NP$ upper bound for $\GMI$. Given
two graphs, $X_1$ and $X_2$, the $\NP$ machine just guesses the
sequence of $n-2$ separating pairs which corresponding to the
2-isomorphism. For each pair, guess the cut w.r.t which the twist
operation is to be done, and apply each of them in sequence to the
graph $X_1$ to obtain a graph $X_1'$. Now ask if $X_1' \cong
X_2'$. This gives an upper bound of $\exists.\GI \subseteq \NP$. Thus we
have,
\begin{proposition}
  \label{prop:npbound}
$\GMI$ is in $\NP$.
\end{proposition}

This can also be seen as an $\NP$-reduction from $\GMI$ to $\GI$.  Now
we will give a deterministic reduction from $\GMI$ to $\GI$. Although,
this does not improve the $\NP$ upper bound, it implies that it is
unlikely that $\GMI$ is hard for $\NP$ (Using methods similar to that
of Proposition~\ref{prop:MIhard}, one can also directly prove that if
$\GMI$ is $\NP$-hard, then $\PH$ collapses to the second level).

\noindent Now we state the main result of the paper:

\begin{theorem}
  \label{thm:gmi-to-gi}
  $\GMI \le_T^p \GI$
\end{theorem}

 Let us first look into the case of 3-connected graphs. A {\em separating
 pair} is a pair of vertices whose deletion leaves the graph
 disconnected. A 3-connected
graph is a connected graph which does not have any separating pairs.
Whitney (\cite{Whi32}) proved the following equivalence,

\begin{theorem}[Whitney, \cite{Whi32}]
\label{thm:3con-GMIGI}
$X_1$ and $X_2$ be 3-connected graphs, 
$X_1 \cong_2 X_2 \iff X_1 \cong X_2$.
\end{theorem}


Before giving a formal proof of Theorem~\ref{thm:gmi-to-gi},
 we describe the idea roughly here:

\paragraph{\bf Basic Idea:} Let $X_1$ and $X_2$ be the given graphs.
From the above discussion, we can assume that the given graph is
2-connected.

In~\cite{HT73}, Hopcroft and Tarjan proved that every 2-connected
graph can be decomposed uniquely into a tree of 3-connected
components, bonds or polygons.\footnote{Cunningham et al. ~\cite{CE80}
shows that any graphic matroid $M(X)$ is isomorphic to $M(X_1)\oplus
M(X_2) \ldots \oplus M(X_k)/\{e_1,e_2, \ldots ,e_k\}$, where $M(X_1),
\ldots,M(X_k)$ are 3-connected components, bonds or polygons of $M(X)$
and $e_1,\ldots, e_k$ are the virtual edges. However, it is unclear if this
can be turned into a reduction from $\GMI$ to $\GI$ using edge/vertex
colouring.}
 Moreover, ~\cite{HT73} showed that this decomposition can
be computed in polynomial time.  The idea is to then find the
isomorphism classes of these 3-connected components using queries to
$\GI$ (see theorem~\ref{thm:3con-GMIGI}), and then colour the tree
nodes with the corresponding isomorphism class, and then compute a
coloured tree isomorphism between the two trees produced from the two
graphs.

A first mind block is that these isomorphisms between the 3-connected
components need not map separating pairs to separating pairs. We
overcome this by colouring the separating pairs (in fact the edge
between them), with a canonical label of the two sub trees which the
corresponding edge connects. To support this, we observe the
following. There may be many isomorphisms between two 3-connected
components which preserves the colours of the separating
pairs. However, the order in which the vertices are mapped within a
separating pair is irrelevant, since any order will be canonical up to
a twist operation with respect to the separating pair.

So with the new colouring, the isomorphism between 3-connected
components maps a separating pair to a separating pair, if and only if
the two pairs of sub trees are isomorphic. However, even if this is the
case, the coloured sub trees need not be isomorphic. This creates a
simultaneity problem of colouring of the 3-connected components and
the tree nodes and thus a second mind block.

We overcome this by colouring again using the code for coloured
sub trees, and then finding the new isomorphism classes between the
3-connected components. This process is iterated till the colours
stabilize on the tree as well as on the individual separating pairs
(since there are only linear number of 3-connected components). Once
this is ensured, we can recover the 2-isomorphism of the original
graph by weaving the isomorphism of the 3-connected components guided
by the tree adjacency relationship. In addition, if two 3-connected
components are indeed isomorphic in the correctly aligned way, the
above colouring scheme, at any point, does not distinguish between
them.

Now we convert this idea into an algorithm and a formal proof. 

\vspace{-2mm} 

\paragraph{\bf Breaking into Tree of 3-connected components:} 

We use the algorithm of Hopcroft and Tarjan \cite{HT73} to compute
the set of 3-connected components of a 2-connected graph in polynomial
time. We will now describe some details of the algorithm which we will
exploit.

Let $X(V,E)$ be a $2$-connected graph. Let $Y$ be a connected
component of $X \setminus \{a, b\}$, where $a,b$ is a separating
pair. $X$ is an {\em excisable} component w.r.t $\{a,b\}$ if $X
\setminus Y$ has at least $2$ edges and is $2$-connected. The
operation of excising $Y$ from $X$ results in two graphs: $C_1 = X
\setminus Y$ plus a {\em virtual edge} joining $(a,b),$ and $C_2 =$
the induced subgraph on $X \cup \{a,b\}$ plus a {\em virtual edge}
joining $(a,b)$. This operation may introduce multiple edges.

The decomposition of $X$ into its 3-connected components is achieved
by the repeated application of the excising operation (we call the
corresponding separating pairs as {\em excised pairs}) until all the
resulting graphs are free of excisable components. This decomposition
is represented by a graph $G_X$ with the 3-connected components of $X$
as its vertices and two components are adjacent in $G_X$ if and only
if they share a virtual edge. In the above explanation, the graph
$G_X$ need not be a tree as the components which share a separating
pair will form a clique.

To make it a tree, \cite{HT73} introduces another component
corresponding to the virtual edges thus identifying all the virtual
edges created in the same excising operation with each other. 

Instead, we do a surgery on the original graph $X$ and the graph
$G_X$. We add an edge between all the {\em excised pairs} (excised
while obtaining $G_X$) to get graph $X'$. Notice that, following the
same series of decomposition gives a new graph $T_X$ which is the same
as $G_X$ except that the cliques are replaced by star centered at a
newly introduced vertex (component) corresponding to the newly
introduced excised edges in $X'$. The newly introduced edges form a
3-connected component themselves with one virtual edge corresponding
to each edge of the clique they replace.

We list down the properties of the tree $T_X$ for further reference.
(1) For every node in $t \in T_X$, there is exactly one 3-connected
component in $X'$. We denote this by $c_t$.~(2) For every edge
$e=(u,v) \in T_X$, there are exactly two virtual edges, one each in
the 3-connected components $c_u$ and $c_v$.  We call these virtual
edges as the {\em twin edges} of each other.~(3) For any given graph
$X$, $T_X$ is unique up to isomorphism (
since $G_X$ is unique~\cite{HT73}). In addition, $T_X$ can be obtained
from $G_X$ in polynomial time.

In the following claim, we prove 
that this
surgery in the graphs does not affect the existence of 2-isomorphisms.

\begin{claim}
  \label{claim:sugery}
$X_1 \cong_2 X_2 \iff X_1' \cong_2 X_2'$.
\end{claim}

\begin{proof}
  Suppose $X_1 \cong_2 X_2$, via a bijection $\phi:E_1 \to E_2$. This
  induces a map $\psi$ between the sets of 3-connected components of
  $X_1$ and $X_2$. By theorem~\ref{thm:3con-GMIGI}, for every
  3-connected component $c$ of $X_1$, $c \cong \psi(c)$ (via say
  $\tau_c$; when $c$ is clear from the context we refer to it as
  $\tau$).

  We claim that $\psi$ is an isomorphism between $G_1$ and $G_2$.  To
  see this, consider an edge $e=(u,v) \in T_1$. This corresponds to
  two 3-connected components $c_u$ and $c_v$ of $X_1$ which share a
  separating pair $s_1$. The 3-connected components $\psi(c_u)$ and
  $\psi(c_v)$ must share a separating pair say $s_2$; otherwise, the
  cycles spanning across $c_u$ and $c_v$ will not be preserved by
  $\phi$ which contradicts the fact that $\phi$ is a
  2-isomorphism. Hence $ (\psi(c_u),\psi(c_v))$ correspond to an edge
  in $G_2$. Therefore, $\psi$ is an isomorphism between $G_1$ and
  $G_2$. In fact, this also gives an isomorphism between $T_1$ and
  $T_2$, which in turn gives a map between the excised pairs of $X_1$
  and $X_2$. To define the 2-isomorphism between $X_1'$ and $X_2'$, we
  extend the map $\psi$ to the excised edges.

  To argue the reverse direction, let $X_1' \cong_2 X_2'$ via
  $\psi$. In a very similar way, this gives an isomorphism between
  $T_1$ and $T_2$. The edge map of this isomorphism gives the map
  between the excised pairs. Restricting $\psi$ to the edges of $X_1$
  gives the required 2-isomorphism between $X_1$ and $X_2$. This is
  because, the cycles of $X_1$($X_2$) are anyway contained in $X_1'$
  ($X_2'$), and the excised pairs does not interfere in the mapping.
\end{proof}

Thus it is sufficient to give an algorithm to test if $X_1' \cong_2 X_2'$,
which we describe as follows.

%
\noindent\line(1,0){470}\\
{\small
{\sc Input:} 2-connected graphs $X_1'$ and $X_2'$ and
  tree of 3-connected components $T_1$ and $T_2$.  \\
  {\sc Output:} {\sc Yes} if $X_1' \cong_2 X_2'$, and {\sc No}
  otherwise.\\ 
  {\sc Algorithm:}\\
  Notation: $\code(T)$ denotes the canonical label\footnote{When $T$
    is coloured, $\code(T)$ is the code of the tree obtained after
    attaching the necessary gadgets to the coloured nodes. Notice that
    even after colouring, the graph is still a tree. In addition, for any
    $T$, $\code(T)$ can be computed in $\L$~\cite{Lin92}.} for a tree $T$.
  \begin{center}
  \begin{enumerate}
    \item Initialize $T_1'=T_1$, $T_2' = T_2$.
    \item {\sc Repeat}
      \label{step:repeat-until}
      \begin{enumerate}
      \item Set $T_1 = T_1'$, $T_2 = T_2'$.
      \item \label{step:treecolor} For each edge $e=(u,v) \in T_i$, $i
        \in \{1,2\}$:

        Let $T_i(e,u)$ and $T_i(e,v)$ be subtrees of $T_i$ obtained by
        deleting the edge $e$, containing $u$ and $v$ respectively.

        Colour virtual edges corresponding to the separating pairs in
        the components $c_u$ and $c_v$ with the set
        $\{\code(T_i(e,u)), \code(T_i(e,v))\}$. From now on, $c_t$
        denotes the coloured 3-connected component corresponding to
        node $t \in T_1 \cup T_2$.
      \item Let $S_1$ and $S_2$ be the set of coloured 3-connected
        components of $X_1'$ and $X_2'$ and let $S = S_1 \cup
        S_2$. Using queries to $\GI$ (see
        observation~\ref{obs:3con-coloring}) find out the isomorphism
        classes in $S$.  Let $C_1, \ldots, C_q$ denote the isomorphism
        classes.
    \item Colour each node $t \in T_i$, $i \in \{1,2\}$, with colour
      $\ell$ if $c_t \in C_\ell$. (This gives two coloured trees
      $T_1^\prime$ and $T_2^\prime$.)
      \end{enumerate}
      {\sc Until} ($\code(T_i) \ne \code(T_i')$, $\forall i \in \{1,2\}$)

    \item Check if $T_1^\prime \cong T_2^\prime$ preserving the
      colours. Answer {\sc Yes} if $T_1^\prime \cong T_2^\prime$, and
      {\sc No} otherwise.
  \end{enumerate}
\end{center}}
\noindent\line(1,0){470}\\

First we prove that the algorithm terminates in linear number of
iterations of the repeat-until loop. Let $q_i$ denote the number of
isomorphism classes of the set of the coloured 3-connected components
after the $i^{th}$ iteration. We claim that, if the termination
condition is not satisfied, then $|q_i| > |q_{i-1}|$. To see this, suppose
the termination is not satisfied. This means that the coloured tree
$T_1'$ is different from $T_1$. This can happen only when the colour of
a 3-connected component $c_v$, $v \in T_1 \cup T_2$ changes. In
addition, this can only increase the isomorphism classes. Thus $|q_i|
> |q_{i-1}|$.  Since $q$ can be at most $2n$, this shows that the
algorithm exits the loop after at most $2n$ steps.
 
Now we prove the correctness of the algorithm. We follow the
notation described in the algorithm.

\begin{lemma}
\label{lem:correctness}
$X_1' \cong_2 X_2'$. $\iff$ $T_1^\prime \cong T_2^\prime$.
\end{lemma}
\begin{proof}
  \noindent{(\sc $\Rightarrow$)} Suppose $X_1' \cong_2 X_2'$, via a
  bijection $\phi:E_1 \to E_2$. This induces a map $\psi$ between the
  sets of 3-connected components of $X_1'$ and $X_2'$. By
  theorem~\ref{thm:3con-GMIGI}, for every 3-connected component $c$ of
  $X_1'$, $c \cong \psi(c)$ (via say $\tau_c$; when $c$ is clear from
  the context we refer to it as $\tau$).

  We claim that $\psi$ is an isomorphism between $T_1$ and $T_2$.
   To see this, consider an edge $e=(u,v) \in T_1$. This corresponds
   to two 3-connected components $c_u$ and $c_v$ of $X_1'$ which share
   a separating pair $s_1$. The 3-connected components $\psi(c_u)$ and
   $\psi(c_v)$ must share a separating pair say $s_2$; otherwise, the
   cycles spanning across $c_u$ and $c_v$ will not be preserved by
   $\phi$ which contradicts the fact that $\phi$ is a
   2-isomorphism. Hence $ (\psi(c_u),\psi(c_v))$ correspond to an edge
   in $T_2$. Therefore, $\psi$ is an isomorphism between $T_1$ and
   $T_2$. So in what follows, we interchangeably use $\psi$ to be a map
   between the set of 3-connected components as well as between
   the vertices of the tree. Note that $\psi$ also induces (and hence
   denotes) a map between the edges of $T_1$ and $T_2$.
   
   Now we prove that $\psi$ preserves the colours attached to $T_1$ and
   $T_2$ after all iterations of the repeat-until loop in
   step~\ref{step:repeat-until}.
   To simplify the argument, we do it for the first iteration and the
   same can be carried forward for any number of iterations.  Let
   $T_1'$ and $T_2'$ be the coloured trees obtained after the first
   iteration. We argue that $\psi$ itself is an isomorphism between
   $T_1'$ and $T_2'$.

   To this end, we prove that for any vertex $u$ in $T_1$, $c_u \cong
   \psi(c_u)$ even after colouring as in step~\ref{step:treecolor}.
   That is, the map preserves the colouring of the virtual edges in
   step~\ref{step:treecolor}.

   Consider any virtual edge $f_1$ in $c_u$, we know that $f_2 =
   \tau(f_1)$ is a virtual edge in $\psi(c_u)$. Let $e_1=(u_1,v_1)$
   and $e_2=(u_2,v_2)$ be the tree edges in $T_1$ and $T_2$
   corresponding to $f_1$ and $f_2$ respectively. We know that, $e_1 =
   \psi(e_2)$. Since $T_1 \cong T_2$ via $\psi$, we have
   $$ \left\{\code(T_1(e_1,u_1)),\code(T_1(e_1,v_1)) \right\} 
   = \left\{\code(T_2(e_2,u_2)),\code(T_2(e_2,v_2)) \right\}.$$
   Thus, in Step~\ref{step:treecolor}, the virtual edges $f_1$ and
   $f_2$ get the same colour. Therefore, $c_u$ and $\psi(c_u)$ belong
   to the same colour class after step~\ref{step:treecolor}. Hence
   $\psi$ is an isomorphism between $T_1'$ and $T_2'$.

   \noindent{(\sc $\Leftarrow$)} First, we recall some definitions
   needed in the proof. A {\em center} of a tree $T$ is defined as a
   vertex $v$ such that $\max_{u \in T} d(u,v)$ is minimized at $v$,
   where $d(u,v)$ is the number of edges in the unique path from $u$
   to $v$. It is known~\cite{Har} that every tree $T$ has a center
   consisting of a single vertex or a pair of adjacent vertices.  The
   minimum achieved at the center is called the {\em height} of the
   tree, denoted by $ht(T)$.

  \begin{claim}
    \label{claim:correctness}
    Let $\psi$ be a colour preserving isomorphism between $T_1'$ and
    $T_2'$, and $\chi_t$ is an isomorphism between the 3-connected
    components $c_t$ and $c_{\psi(t)}$. Then, $X_1' \cong_2 X_2'$ via
    a map $\sigma$ such that $\forall t \in T_1'$, $\forall e \in c_t
    \cap E_1 : \sigma(e) = \chi_t(e)$ where $E_1$ is the set of edges
    in $X_1'$.
  \end{claim}
  \begin{proof}
  The proof is by induction on height of the trees $h = ht(T_1^\prime)
  = ht(T_2^\prime)$, where the height (and center) is computed with
  respect to the underlying tree ignoring colours on the vertices.

  Base case is when $h=0$; that is, $T_1^\prime$ and $T_2^\prime$ 
  have just one node (3-connected component) 
  without any virtual edges. Simply define $\sigma = \chi$. By
  Theorem~\ref{thm:3con-GMIGI}, this gives the required
  2-isomorphism.
%

  Suppose that if $h = ht(T_1^\prime) = ht(T_2^\prime) < k$, the above
  claim is true. For the induction step, suppose further that $T_1^\prime
  \cong T_2^\prime$ via $\psi$, and $ht(T_1^\prime) = ht(T_2^\prime) =
  k$. Notice that $\psi$ should map the center(s) of $T_1$ to that of
  $T_2$. We consider two cases (we present one case here, and the
  other in the appendix).

  In the first case, $T_1^\prime$ and $T_2^\prime$ have unique centers
  $\alpha$ and $\beta$. It is clear that $\psi(\alpha) = \beta$. Let
  $c_1$ and $c_2$ be the corresponding coloured (as in
  step~\ref{step:treecolor}) 3-connected components. Therefore, there
  is a colour preserving isomorphism $\chi = \chi_{\alpha}$ between
  $c_{\alpha}$ and $c_{\beta}$.  Let $f_1, \ldots f_k$ be the virtual
  edges in $c_{\alpha}$ corresponding to the tree edges
  $e_1=(\alpha,v_1), \ldots, e_k=(\alpha,v_k)$ where $v_1, \ldots,
  v_k$ are neighbors of $\alpha$ in $T_1^\prime$. Denote $\psi(e_i)$
  by $e_i'$, and $\psi(v_i)$ by $v_i'$.

  Observe that only virtual edges are coloured in the 3-connected
  components in step~\ref{step:treecolor} while determining their
  isomorphism classes.  Therefore, for each $i$, $\chi(f_i)$ will be a
  virtual edge in $c_{\beta}$, and in addition, with the same colour
  as $f_i$. That is,
  \begin{center}
  {\small
  $\left\{ \code(T_1(e_i,\alpha)),\code(T_1(e_i,v_i)) \right\} =
  \left\{ \code(T_2(e_i',\beta)),\code(T_2(e_i',v_i'))) \right\}$.}
  \end{center}
  Since $\alpha$ and $\beta$ are the centers of $T_1'$ and $T_2'$, it
  must be the case that in the above set equality,
  $\code(T_1(e_i,v_i))$ $=$ $\code(T_2(e_i',v_i'))$. From the termination
  condition of the algorithm, this implies that $\code(T_1'(e_i,v_i))
  = \code(T_2'(e_i',v_i'))$. Hence, $T_1'(e_i,v_i) \cong
  T_2'(e_i',v_i')$. In addition, $ht(v_i)=ht(v_i') < k$.  Let
  $X_{f_i}'$ and $X_{\chi(f_i)}'$ denote the subgraphs of $X_1'$ and
  $X_2'$ corresponding to $T_1'(e_i,v_i)$ and $T_2'(e_i',v_i')$
  respectively. By induction hypothesis, the graphs $X_{f_i}'$ and
  $X_{\chi(f_i)}'$ are 2-isomorphic via $\sigma_i$ which agrees with
  the corresponding $\chi_t$ for $t \in T_1'(e_i,v_i)$. Define
  $\pi_i$ as a map between the set of all edges, such that it agrees
  with $\sigma_i$ on all edges of $X_{f(i)}'$ and with $\chi_t$ (for
  $t \in T_1'(e_i,v_i)$) on the coloured virtual edges.

  We claim that $\pi_i$ must map the twin-edge of $f_i$ to twin-edge
  of $\tau(f_i)$. Suppose not. By the property of the colouring, this
  implies that there is a subtree of $T_1'(e_i,v_i)$ isomorphic to
  $T_1^\prime \setminus T_1'(e_i,v_i)$. This contradicts the
  assumption that $c_{\alpha}$ is the center of $T_1^\prime$.

  For each edge $e \in E_1$, define $\sigma(e)$ to be $\chi(e)$ when $e \in c_{\alpha}$
  and to be $\pi_i(e)$ when $e \in E_{f_i}~(\textrm{edges of $X_{f_i}$})$.
  From the above argument, $\chi = \chi_{\alpha}$ and $\sigma_i$
  indeed agrees on where it maps $f_i$ to. This ensures that every
  cycle passing through the separating pairs of $c_{\alpha}$ gets
  preserved. Thus $\sigma$ is a 2-isomorphism between $X_1'$ and
  $X_2'$. 

  For case 2, let $T_1^\prime$ and $T_2^\prime$ have two centers
  $(\alpha_1,\alpha_2)$ and $(\beta_1,\beta_2)$ respectively. It is
  clear that $\psi(\{\alpha_1,\alpha_2\}) =
  \{\beta_1,\beta_2\}$. Without loss of generality, we assume that
  $\psi(\alpha_1) = \beta_1$, $\psi(\alpha_2) = \beta_2$. Therefore,
  there are colour preserving isomorphisms $\chi_1$ from
  $c_{\alpha_1}$ to $c_{\beta_1}$ and $\chi_2$ from $c_{\alpha_2}$ and
  $c_{\beta_2}$. Define $\chi(e)$ as follows:
    \[
    \chi(e) = \left\{
    \begin{array}{ll}
       \chi_1(e) & ~~e \in c_{\alpha_1} \\
       \chi_2(e) & ~~e \in c_{\alpha_2}
    \end{array}
    \right.
    \]
    \[ c_{\alpha} = \cup_{i} c_{\alpha_i}, ~~~ c_{\beta} = \cup_{i} c_{\beta_i} \]

    With this notation, we can appeal to the proof in the case 1, and
    construct the 2-isomorphism $\sigma$ between $X_1'$ and $X_2'$.

\end{proof}
This completes the proof of correctness of the algorithm (Lemma
~\ref{lem:correctness}).

\end{proof}

To complete the proof of Theorem~\ref{thm:gmi-to-gi}, we need the
following observation,

\begin{observation}
  \label{obs:3con-coloring}
  $\cGMI$ for 3-connected graphs reduces to $\GI$.
\end{observation}

Observing that the above construction does not use non-planar gadgets,
we get the following.

\begin{corollary}
\label{cor:planarMI}
Given two planar matroids, $M(X_1)$ and $M(X_2)$, testing if
$M(X_1) \cong M(X_2)$ can be performed in $\P$.
\end{corollary}


Now we give a polynomial time many-one reduction from $\MI_b$ to $
\GMI$. Let $M_1$ and $M_2$ be two matroids of rank $b$ over the ground
set $S_1$ and $S_2$. Let ${\mathcal C}_1$ and ${\mathcal C}_2$ respectively
denote the set of cycles of $M_1$ and $M_2$. Note that $|{\mathcal C}_1|,
|{\mathcal C}_2| \le m^{b+1}$.

Define graphs $X_1=(V_1,E_1)$ (respectively for $X_2=(V_2,E_2)$) as
follows. For each circuit $c =\{s_{i_1},\ldots ,s_{i_\ell}\} \subseteq
S_1$ in $M_1$, $X_1$ contains a simple cycle $\{e(c,s_{i_1}), \ldots ,
e(c,s_{i_\ell}) \}$. Now pairwise interconnect all the endpoints of the edges
corresponding to each of the ground set elements (these edges  form a
clique), and colour these edges as {\sc red} and the remaining edges
as {\sc blue}. Now we claim the following.

\begin{lemma}
  \label{lem:migmi}
  $M_1 \cong M_2 $ if and only if $X_1 \cong_2 X_2$.
\end{lemma}

\begin{proof}
  Suppose $M_1 \cong M_2$, via a map $\phi :S_1 \to S_2$. 
  This gives a map $\psi$ between the {\sc blue} edges of the graphs
  $X_1$ and $X_2$ which preserves {\sc blue} cycles. Now we extend this
  to the {\sc red} edges. Take a {\sc red} edge $r$, there are two {\sc blue}
  edges $e_1$ and $e_2$ which share an endpoint with $r$. We know that
  $e_1$ and $e_2$ are corresponding to the same ground set element
  (say $s$). Thus $\psi(e_1)$ and $\psi(e_2)$ correspond to the same
  ground set element $\phi(s)$, and hence shares a {\sc red} edge in
  $X_2$.  Thus $\psi$ can be extended to preserve the {\sc red}
  edges. Hence $X_1 \cong_2 X_2$.

  Conversely, suppose $X_1 \cong_2 X_2$ via $\psi : E_1 \to E_2$.
  Define $\phi: S_1 \to S_2$ as follows: For $s \in S_1$ let $R_s$
  denote the clique in $X_1$ corresponding to $s$. $R_s$ is either a
  single blue edge or a clique on at least $4$ vertices (in the latter
  case it is 3-connected). Thus $\psi$ should map $R_s$ to $R'_{s'}$
  for some $s'$ in $S_2$..  Define $\phi(s) = s'$.
  Now we argue that $\psi$ is an isomorphism between $M_1$ and $M_2$.
  Let $c=\{s_1,\ldots, s_\ell\} \subseteq S_1$ be a cycle in
  $M_1$.
  \begin{eqnarray*}
    c \in {\mathcal C}_1 & \iff & \bigcap_i\psi(R_{s_i}) 
    \textrm{ is a {\sc blue} cycle in $X_1$} \\ 
    & \iff & \bigcap_i\psi(R'_{s'_i})
    \textrm{ is a {\sc blue} cycle in $X_2$} \\
    & \iff & \phi(c) \in {\mathcal C}_2
  \end{eqnarray*}
\end{proof}

From the above construction, we have the following theorem.
\begin{theorem}
  \label{thm:bmi-gmi}
  $\MI_b \le^p_m \GMI$.
\end{theorem}
Thus we have,
\begin{theorem}
\label{thm:gmi-gi-equiv}
$\GI \equiv ^p_T \GMI\equiv^p_T \MI_b \equiv^p_T \LMI_b $
\end{theorem}

\section{Matroid Automorphism Problem}
With any isomorphism problem, there is an associated automorphism
problem i.e, to find a generating set for the automorphism group of
the underlying object. Relating the isomorphism problem to the
corresponding automorphism problem gives access to algebraic tools
associated with the automorphism groups. In the case of graphs, studying automorphism
problem has been fruitful.(e.g. see
\cite{Luk80,BGM82,AK02}.) In this section we turn our attention to
Matroid automorphism problem.

An automorphism of a matroid $M = (S,\C)$ (where $S$ is the ground set
and $\C$ is the set of circuits) is a permutation $\phi$ of elements of $S$
such that $\forall C \subseteq S,~ C \in \C \iff \phi(C) \in
\C$. $Aut(M)$ denotes the group of automorphisms of the matroid
$M$. When the matroid is graphic we denote by $Aut(X)$ and $Aut(M_X)$
the automorphism group of the graph and the graphic matroid
respectively.

To begin with, we note that given a graph $X$, and a permutation $\pi
\in S_m$, it is not clear apriori how to check if $\pi \in Aut(M_X)$
efficiently. This is because we need to ensure that $\pi$ preserves
all the simple cycles, and there could be exponentially many of
them. Note that such a membership test (given a $\pi \in S_n$) for
$Aut(X)$ can be done easily by testing whether $\pi$ preserves all the
edges. We provide an efficient test for this problem. 

We use the notion of a cycle bases of $X$. A {\em cycle basis} of a
graph $X$ is a minimal set of cycles $\mathcal{B}$ of $X$ such that every
cycle in $X$ can be written as a linear combination (viewing every
cycle as a vector in $\F_2^m$) of the cycles in $\B$. Let
$\mathscr{B}$ denote the set of all cycle basis of the graph $X$.

\begin{lemma}
\label{lem:cb-exists-forall} 
Let $\pi \in S_n$, $\exists \B \in \BB: \pi(\B) \in \BB 
\implies \forall \B \in \BB:\pi(\B) \in \BB $
\end{lemma}

\begin{proof}
  Let $\B = \{b_1, \ldots b_\ell \} \in \BB$ such that $\pi(\B) =
  \{\pi(b_1), \ldots, \pi(b_\ell) \}$ is a cycle basis. Now consider
  any other cycle basis $\B'= \{ b_1', \ldots, b_k' \} \in \BB$. Thus,
  $b_i = \sum_{j} \alpha_j b_j'$. This implies,
  \[ \pi(b_i) = \sum_{j} \alpha_j \pi(b_j'). \] Thus, $\pi(B') =
  \{\pi(b_1'),\ldots,\pi(b_\ell') \}$ forms a cycle basis.
\end{proof}

\begin{lemma}
\label{lem:aut-cb}
Let $\pi \in S_m$, and let $\B \in \BB$, then
$\pi \in Aut(M_X) \iff \pi(\B) \in \BB $.
\end{lemma}

\begin{proof}
  Let $\B = \{b_1, \ldots, b_\ell\}$ be the given cycle basis. 

  For the forward direction, suppose $\pi \in Aut(M_X)$. That is, $C
  \subseteq E$ is a cycle in $X$ if and only if $\pi(C)$ is also a
  cycle in $X$.  Let $C$ be any cycle in $X$, and let
  $D=\pi^{-1}(C)$. Since $\B \in \BB$, we can write, $D = \sum_i
  \alpha_i b_i$, and hence $C = \sum _i \alpha_i \pi(b_i)$.  Hence
  $\pi({\mathcal B})$ forms a cycle basis for $X$.

  For the reverse direction, suppose $\pi({\mathcal B})$ is a cycle basis
  of $X$.
 Let $C$ be any cycle in
  $X$. We can write $C=\sum_i\alpha_i b_i$. 
   Hence, $\pi(C) =\sum_i \alpha_i
  \pi(b_i)$. As $\pi$ is a bijection, we have $\pi(b_i\cap b_j )=
  \pi(b_i) \cap \pi(b_j) $. Thus $\pi(C)$ is also a cycle in
  $X$. Since $\pi$ extends to a permutation on the set of cycles, we
  get that $C$ is a cycle if and only if $\pi(C)$ is a cycle.
\end{proof}

Using Lemmas~\ref{lem:cb-exists-forall} and ~\ref{lem:aut-cb}
it follows that, given a permutation $\pi$, to
test if $\pi \in Aut(M_X)$ it suffices to check if for a cycle basis
$\B$ of $X$, $\pi(\B)$ is also a cycle basis. Given a graph $X$ a
cycle basis ${\mathcal{B}}$ can be computed in polynomial time (see e.g,
\cite{Hor87}). Now it suffices to show:

\begin{lemma}
  Given a permutation $\pi \in S_m$, and a cycle basis $\B \in \BB$,
  testing whether $\pi(\B)$ is a cycle basis, can be done in
  polynomial time.
\end{lemma}
\begin{proof}
  To check if $\pi(\B)$ is a cycle basis, it is sufficient to verify
  that every cycle in $\B = \{b_1, \ldots, b_\ell\}$ can be written as
  a $\F_2$-linear combination of the cycles in $\B'= \{b_1', \ldots,
  b_\ell'\} = \pi(\B)$. This can be done as follows.

  For $b_i \in \B$, let $\pi(b_i)=b_i'$. View $b_i$ and $b_i'$ as
  vectors in $\F_2^m$. Let $b_{ij}$ (resp. $b_{ij}'$) denote the
  $j^{\mathrm{th}}$ component of $b_i$ (resp. $b_i'$). Construct the
  set of linear equations, $b_{ij}' = \sum_{b_k \in \B} x_{ik} b_{kj}$
  where $x_{ik}$ are unknowns. There are exactly $\ell$ $b_i'$s and
  each of them gives rise to exactly $m$ equations like this. This
  gives a system $I$ of $\ell m$ linear equations in $\ell^2$ unknowns
  such that, $\pi(B)$ is a cycle basis if and only if $I$ has a
  non-trivial solution. This test can indeed be done in polynomial
  time.
\end{proof}

This gives us the following:
\begin{theorem}
  Given any $\pi \in S_m$, the membership test for $\pi$ in $Aut(M_X)$
  is in $\P$.
\end{theorem}

Notice  that  similar  arguments   can  also  give  another  proof  of
Proposition~\ref{prop:npbound}. As in the case of graphs, we
can 
define automorphism problems for
matroids. 

{\em $\ma(\MA)$: Given a matroid $M$ as independent set oracle, 
compute a generating set for $Aut(M)$.}

We define $\GMA$ and $\LMA$ 
as the corresponding
automorphism problems for graphic and linear matroids,  when the input
is a graph and matrix respectively. 
Using the colouring techniques from Section~\ref{sec:coloring}, we
prove the following equivalence.

\begin{theorem}
\label{thm:iso-auto}
$\LMI \equiv_T^p \LMA$, and 
$\GMI \equiv_T^p \GMA$.
\end{theorem}

\begin{proof}
  This proof follows a standard idea due to Luks~\cite{Luks}. We argue
  the forward direction as follows. Given two matrices $M_1$ and
  $M_2$, form the new matrix $M$ as,
  \[ M = \begin{bmatrix} M_1 & 0 \\ 0 & M_2 \end{bmatrix} \] 
  Now using queries to $\LMA$ construct the generating set of
  $Aut(M)$.  Check if at least one of the generators map the columns in
  $M$ corresponding to columns of $M_1$ to those corresponding to the
  columns of $M_2$.

  To see the other direction, we use the colouring idea, and the rest
  of the details is standard. The idea is to find the orbits of each
  element of the ground set as follows: For every element of $e \in
  S$, for each $f \in S$, colour $e$ and $f$ by the same colour to obtain
  coloured matroids $M_1$ and $M_2$.  Now by querying to $\LMI$ we can
  check if $f$ is in the orbit of $e$. Thus we can construct the orbit
  structure of $Aut(M)$ and hence compute a generating set.
  
  Using similar methods we can prove $\GMI \equiv_T^p \GMA$.
\end{proof}

\section{Conclusion and Open Problems}

We studied the matroid isomorphism problem under various input
representations and restriction on the rank of the matroid. We proved
that graph isomorphism,
graphic matroid isomorphism and  bounded rank version of matroid isomorphism
are all polynomial time equivalent.

In addition, we find it interesting that in the bounded rank case,
$\MI_b$ and $\LMI_b$ are equivalent, though there exist matroids of
bounded rank which are not representable over any field.  Some of the
open questions that we see are as follows:

\begin{itemize}
\item Our results provide new possibilities to attack the graph
  isomorphism problem. For example, it will be interesting to prove a
  $\co\NP$ upper bound for $\LMI_b$. Note that this will imply that
  $\GI \in \NP \cap \co\NP$. Similarly, are there special cases of
  $\GMI$ (other than what is translated from the bounds for $\GI$)
  which can be solved in polynomial time?

\item The representations of the matroid in the definition of $\LMI$
  is over fields of size at least $m$ and at most $\poly(m)$, where
  $m$ is the size of the ground set of the matroid. This is critically
  needed for the observation of $\co\NP$-hardness. One could ask if
  the problem is easier over fixed finite fields independent on the
  input. However, we note that, by our results, it follows that this
  problem over $\F_2$ is already hard for $\GI$. It will still be
  interesting to give a better (than the trivial $\Sigma_2$) upper
  bound for linear matroids represented over fixed finite fields (even
  for $\F_2$).
\item Can we use the colouring technique of linear matroid isomorphism
  to reduce the general instances of linear matroid isomorphism to
  isomorphism testing of ``simpler components'' of the matroid?
\item Can we make the reduction from $\GMI$ to $\GI$ many-one? Can we
 improve the complexity of this reduction in the general case?

\end{itemize}

\section{Acknowledgements}
We thank V.~Arvind and Meena Mahajan for providing us with useful
inputs and many insightful discussions, James Oxley for sharing his
thoughts while responding to our queries about matroid isomorphism.
We also thank the anonymous referees for providing us many useful
pointers to the literature.

\bibliographystyle{alpha}
\bibliography{matroids}

\newpage
\appendix

\section{Proof of Claim~\ref{claim:babai}}
  \begin{aclaim}{\ref{claim:babai}}\cite{Bab78}
    For any graph $X$, any hyperplane of $St_k(X)$ is a maximal set
    edges which share a common vertex (forms a star) in $X$, and these
    are the only hyperplanes.
  \end{aclaim}
  \begin{proof}
  To see the first part, note that since $k$ is at most the min-degree
  of the graph, any maximal set of edges which forms a star is a
  hyperplane. To argue that these are the only hyperplanes, suppose
  $H$ is a hyperplane whose edges do not share a common vertex in
  $X$. Since any independent set cannot be a hyperplane, $H$ has size
  at least $k$, and the rank of $H$ by definition is $k-1$. Take any
  $k$ sized subset $S$ of $H$. It has to be a star (say at vertex
  $v$), since otherwise $H$ will have rank $k$. Since $H$ by itself is
  not a star, there is an edge $e$ which does not have $v$ as its end
  point. But then, $H$ contains a $k$ sized subset which is not a
  star, and hence independent. This is a contradiction.
  \end{proof}

\section{3-connected $\cGMI$}
By combining Lemma~\ref{lem:coloring} with
Theorem~\ref{thm:3con-GMIGI}, we prove the following corollary which
we present although we do not need it explicitly in the paper.

\begin{corollary}
  \label{cor:3con-coloring}
  Let $X_1$ and $X_2$ be two 3-connected graphs with given edge
  colourings, testing if there is a colour preserving 2-isomorphism
  between $X_1$ and $X_2$, can be reduced to $\GI$.
\end{corollary}
\begin{proof}
We follow the notation in the proof of Lemma~\ref{lem:coloring}.  Let
$X_1'$ and $X_2'$ be the two graphs obtained from $X_1$ and $X_2$ by
attaching the colouring gadgets. Now we claim that $X_1' \cong_2 X_2' \iff
X_1' \cong X_2'$.

One direction is easy to see. $X_1' \cong X_2'  \Rightarrow X_1' \cong_2
X_2'$. To see the other direction, suppose $X_1' \cong_2 X_2'$ via
$\sigma$. By proof of Lemma~\ref{lem:coloring} this induces a
2-isomorphism between $X_1$ and $X_2$, which in turn induces an
isomorphism $\tau$ between $X_1$ and $X_2$. Observe that $\tau$ also
preserves colours on the edges of $X_1$ and $X_2$.

Let $\pi$ be the map between the vertices of $X_1'$ and $X_2'$ induced
by $\tau$. Note that $\pi$ is not defined on the new vertices
introduced by the colouring gadgets. However, $\pi$ already preserves
all the edges between $X_1'$ and $X_2'$ except the newly introduced edges.

Since $\tau$ preserves edge colours in $X_1$, any coloured edge $e \in
X_1$, the paths $P_e$ and $P_{\tau(e)}$, introduced while constructing
$X_1'$ and $X_2'$ are of the same length. Hence the vertex map $\pi$
can be extended to the vertices in these paths to an edge preserving
vertex map between $X_1'$ and $X_2'$.
\end{proof}

\section{A hardness result for $\GMI$}

We show that known hardness results of $\GI$ (\cite{Tor04}) can be
adapted to the case of $\GMI$. This is subsumed by the many-one
reduction from $\GI$ to $\GMI$. But we state this observation here
anyway.

\begin{theorem}
\label{thm:NL-hard}
$\GMI$ is $\NL$-hard under $\AC^0$ many one reductions.
\end{theorem}

The hardness is proved using ideas similar to that in \cite{Tor04}.
We include the details of this modified part of the proof.

Consider the graph $X(k) = (V,E)$ where, 
\begin{eqnarray*}
V & = & \{ x_0, \ldots, x_{k-1},y_0, \ldots, y_{k-1}, z_0, \ldots , z_{k-1} \} \\
& & \cup \{u_{i,j} ~|~ 0 \le i,j \le k-1 \} \\
E & = & \{ (x_i,u_{i,j}) ~|~ 0 \le i,j \le k-1 \} \\
& & \cup \{ (y_j,u_{i,j})~|~ 0 \le i,j \le k-1 \} \\ 
& & \cup \{(u_{i,j},z_{i\oplus j})~|~ 0 \le i,j \le k-1 \}
\end{eqnarray*}

The following is easy to verify for the above graph.
\begin{observation}
For $k \ge 3$, the graph $X(k)$ is 3-connected.
\end{observation}

Toran \cite{Tor04} argued that for any $a,b \in \{0, \ldots, k-1\}$
the graph $X(k)$ has a unique automorphism which maps vertices $x_i
\to x_{a \oplus i}$ and $y_i \to y_{b \oplus i}$, which also maps $z_i
\to z_{i\oplus a \oplus b}$. Combining this with Proposition
\ref{thm:3con-GMIGI}, and the above observation, we get the
following lemma.

\begin{lemma}
For any $k \ge 3$, given any $a,b \in \{0, \ldots, k-1\}$, there
exists a unique automorphism for the matroid $M_{X(k)}$ which maps
\begin{eqnarray*}
(x_i,u_{i,j}) & \to & (x_{i \oplus a}, u_{(i \oplus a, j)}) \\ 
(y_j,u_{i,j}) & \to & (y_{i \oplus b}, u_{(i, j \oplus a)}) \\
(u_{i,j}, z_{i \oplus j}) & \to & (u_{i \oplus a, j \oplus b}, 
z_{i \oplus j \oplus a \oplus b}) 
\end{eqnarray*}
\end{lemma}

\begin{lemma}
For $k \ge 3$, $\GMI$ is $\Mod_k\L$-hard under $\AC^0$ many-one
reductions.
\end{lemma}

\begin{proof}
Let $C$ be a circuit with $\Mod_k$ gates for $k \ge 3$. Without loss
of generality, we assume that each gate is of fan-in 2. We construct a
new graph $X$ as follows : For each gate $g$ in $C$, $X$ contains a
copy of the graph $X(k)$, denoted by $X_g$. Colour each edge of $X_g$
with colour $g$ (see Lemma \ref{lem:coloring}). If gate $g$ has $h_1$
and $h_2$ as its inputs identify $x_0, \ldots x_{k-1}$ and $y_0,
\ldots, y_{k-1}$ of $X_g$ with $z_0, \ldots, z_{k-1}$ of $X_{h_1}$ and
$X_{h_2}$ respectively. Let $r$ be the root gate of $C$, $g_1, \ldots
g_m$ be the gates which receive the inputs directly, then the
following claim can be verified by induction on the circuit structure.

\begin{claim}
$C$ evaluates to $\ell \in \{0, \ldots, k-1\}$ on input $a_1, \ldots,
  a_n$ if and only if $M_X$ has an automorphism which maps, for the
  gate $g_i$ receiving the input $a_j$ as its 'x' input, in $X_{g_i}$
\[ (x_i,u_{i,t}) \to (x_{i \oplus a_j}, u_{(i \oplus a_j, t)}) \] 
for all $0 \le t \le k-1$, and in $X_{r}$, for all $a \oplus b =
\ell$, $j \oplus t = i$,
\[ (u_{t,j}, z_i) \to (u_{t \oplus a, j \oplus b}, z_{i \oplus \ell}) \]
\end{claim}
\end{proof}

\begin{proof}(of theorem \ref{thm:NL-hard})
Using ideas from \cite{Tor04}, it is easy to see that we can write an
$\NL$ computation as a series $\Mod_k\L$ computations, for $3 \le k \le
2n$. Thus, the theorem follows.
\end{proof}

The following corollary is immediate as in (Theorem 4.4,
\cite{Tor04}), using Chinese remaindering, and the above reduction.
\begin{corollary}
Every $\Gap\L$-function is $\AC^0$ many-one reducible to $\GMI$.
\end{corollary}

\section{Testing Uniformity of Matroids}

A problem of testing if a given matroid is uniform, is clearly as a
special case of the matroid isomorphism problem, where one of the
matroid is uniform. The problem is known to be $\co\NP$-complete as
shown below.

\begin{proposition}
  Given a representable matroid of rank $k$ (input is a $k \times n$
  matrix $A$), the problem of checking if the matroid is uniform is
  $\co\NP$-complete. In other words, testing if the matroid $M$
  represented by columns of $A$ is isomorphic to $U_{k,n}$ is
  $\co\NP$-complete.
\end{proposition}

\begin{proof}
  This proof is due to Oxley and Welsh \cite{OW02}. We make it more
  explicit.  A set of $n$ points in $d$ dimensions is {\em linearly
    degenerate} if there is a set of $d$ points which is linearly
  dependent. In other words, the set of $n$ points is said to be in
  {\em general position} if all subsets of size $d$ are linearly
  independent.  Khanchiyan \cite{Kha95} 
  proved that given $n$ points
  testing if they are in general position is $\NP$-hard. Now notice
  that this exactly answers the question of testing if the matroid 
  is uniform.
\end{proof}
\end{document}